\documentclass[a4paper,11pt]{article}
\pdfoutput=1 
\usepackage{jheppub} 
\usepackage{tikz}
\usepackage[T1]{fontenc} 
\usepackage{mciteplus}
\usepackage{hyperref}
\usepackage{amssymb}
\usepackage{amsthm}
\usepackage{amstext}
\usepackage{amsmath}
\usepackage{amsfonts}
\usepackage{bbm}
\usepackage{bm} 
\usepackage[german,english]{babel}

\usepackage[colorinlistoftodos]{todonotes}

\usepackage{xcolor}

\usepackage{enumitem}
\usepackage{algorithm}
\usepackage{algorithmic}
\theoremstyle{plain}

\theoremstyle{definition}

\newcommand*{\diff}{\mathop{}\!\mathrm{d}}
\newcommand{\eps}{\epsilon}
\newcommand*{\bz}{\bm{z}}

\title{\boldmath Leading singularities in Baikov representation and Feynman integrals with uniform transcendental weight}

\author[a]{Christoph Dlapa}
\author[b]{Xiaodi Li}
\author[c,d]{and Yang Zhang}

\affiliation[a]{Max-Planck Institute Physics, Munich, 80805, Germany}
\affiliation[b]{Department of Physics, Zhejiang University, China}
\affiliation[c]{Interdisciplinary Center for Theoretical Study, University of Science and Technology of China, Hefei, Anhui 230026, China}
\affiliation[d]{Peng Huanwu Center for Fundamental Theory, Hefei, Anhui 230026, China}

\emailAdd{dlapa@mpp.mpg.de}
\emailAdd{lixiaodi@zju.edu.cn}
\emailAdd{yzhphy@ustc.edu.cn}

\preprint{MPP-2021-23, PCFT-21-10, USTC-ICTS-21-10}

\abstract{We provide a leading singularity analysis protocol in Baikov
representation, for the searching of Feynman integrals with uniform
transcendental (UT) weight. This approach is powered by the recent developments
in rationalizing square roots and syzygy computations, and is
particularly suitable for finding UT integrals with multiple mass scales. We demonstrate the
power of our approach by determining the UT basis for a two-loop
diagram with three external mass scales.}

\begin{document} 
\maketitle
\flushbottom

\section{Introduction}
\label{sec:intro}
After the Run II of the Large Hadron Collider (LHC), high energy physics enters the era of precision physics. To reduce the theoretical uncertainty in high energy physics, it is necessary to calculate the Standard Model (SM) and SM-like models to high orders precisely in the perturbation theory of quantum fields. A hardcore problem of perturbation theory, is the computation of Feynman integrals of multi-loop order, or/and with multiple kinematic scales. 

There are many different ways of computing Feynman integrals \cite{Smirnov:2012gma}, via integral parameterizations \cite{Bergere:1973fq, Usyukina:1975yg, Baikov:1996cd}, sector decomposition \cite{Hepp:1966eg, Speer:1975dc, Heinrich:2008si, Borowka:2017idc, Borowka:2018goh}, dimension recursion relations \cite{Laporta:2001dd,Lee:2009dh}, and differential equation \cite{Kotikov:2000ye,Remiddi:1997ny}. The canonical differential equation approach, with Feynman integrals with uniform transcendental weight \cite{Henn:2013pwa}, is a milestone of the computation of Feynman integrals. By this method, the canonical differential equation for Feynman integrals has a simple form
\begin{equation}
  \label{eq:1}
  d I =\epsilon (d \tilde A) I,
\end{equation}
where $I$ is an integral basis with {\it uniform transcendental (UT) weight}. $\tilde A$ is a matrix where each entry is a sum of logarithms of so-called {\it symbol letters}. Once the canonical differential equation is obtained and rationalized, and the boundary condition is determined, the integral basis can be easily calculated by iterative integrations to an analytic result. The canonical differential equation method has been used widely for the Feynman integral evaluation in Standard Model precision physics and in formal aspects of quantum field theories.

One crucial step in applying the canonical differential equation approach is to find an integral basis with uniform transcendental weight. Significant effort has been put in designing methods and algorithms for finding a UT basis, for example, by the four-dimensional leading singularity analysis \cite{Henn:2013pwa,Henn:2014qga}, by the Magnus series \cite{Argeri:2014qva}, by the {\it dlog} integrand construction \cite{Wasser:2018qvj} with the four-dimensional integrand or the Baikov representation \cite{Chicherin:2018old}, by the initial algorithm \cite{Dlapa:2020cwj}, by the Poincare index computations (Lee's algorithm) \cite{Lee:2014ioa, Lee:2017oca, Lee:2020zfb}, by the intersection theory \cite{Chen:2020uyk, Frellesvig:2020qot} and etc. In recent years, there is a great progress of the UT basis determination, and there are several publicly available packages for determining a UT basis, like {\sc Canonica} \cite{Meyer:2016slj,Meyer:2017joq}, {\sc Fuchsia} \cite{Gituliar:2017vzm}, {\sc epsilon} \cite{Prausa:2017ltv}, {\sc initial} \cite{Dlapa:2020cwj} and {\sc libra} \cite{Lee:2020zfb}.
However, in general, it is still not an easy task to find a UT basis for multi-loop Feynman integrals, especially for the cases with multiple scales.

In this paper, we further develop the Baikov leading singularity analysis method, powered by the new developments in the syzygy based integration-by-parts (IBP) reduction method on the Baikov integrand and modern algebraic methods for rationalizing square roots. Our method can be summarized as the following steps:
\begin{enumerate}
\item In a Feynman integral family, derive the Baikov representation (the usual representation or the loop-by-loop approach) of every sector.
\item In each sector, if necessary, rationalize the Baikov integrand with our rationalization package, to reduce the number of roots in the integrand. 
\item For each sector, with a leading singularity analysis, find the candidate UT integrands. If one candidate UT integrand does not have the form of Feynman integrals in the family, we apply the syzygy IBP reduction to convert it to a ``normal'' Feynman integral's Baikov representation. 
\end{enumerate}
Note that the UT candidates on one sector we found, may not correspond to the master integral counting in the sense of Laporta algorithm. The reason is that we frequently use Laporta-reducible integrals in higher sectors as our UT basis candidates. Once the complete (and independent) candidate UT basis is obtained, we use the IBP reduction (directly or by the finite-field reconstruction) to derive the differential equation, to verify if the canonical differential equation is obtained.

We present a cutting-edge example to demonstrate our method, the two-loop double box integral family with three different external masses. This example is complicated since it contains five kinematic variables, and the symbol letter structure is complicated. The traditional 4D leading singularity analysis or the recent Baikov analysis does not straightforwardly apply on this integral family, since the leading singularities contain complicated roots. However, with our approach, this difficulty is overcome by the modern rationalization tool. We determine the UT basis of this integral family  smoothly, and obtain the analytic canonical differential equation as well as symbol letters.

This paper is organized as follows: In section \ref{Singularity}, we further develop the leading singularity analysis in Baikov representation and the method to find UT integral candidates.
In section \ref{sec:dbox3m2}, in detail, we present the example of the double box integral family with three different external masses, with a sector-by-sector Baikov leading singularity analysis. In section \ref{sec:syz}, we show that in cases needed, we can convert a non-traditional integrand in Baikov representation to the Baikov representation of normal Feynman integrals, by the syzygy IBP method. By this method, we re-derive some of the UT integrals in the three-external-mass double box integral family as examples. In section \ref{sec:summary}, we summarize our method and give an outlook of the further developments.

\section{Singularity analysis in Baikov representation}
\label{Singularity}
\subsection{Leading Singularities and dlog integrals}

Given an $L$-loop Feynman integral,
\begin{align}
\label{eq:Feynman_integral}
I_{a_1,\ldots,a_n}=e^{L \gamma_{E} \epsilon}\int \prod_{k=1}^{L} \frac{\diff^D l_k}{i\pi^{D/2}} \frac{1}{D_1^{a_1}\cdots D_n^{a_n}},\qquad D=4-2\eps,
\end{align}
it has been conjectured\footnote{We explicitly consider integrals that evaluate to multiple polylogarithms and do not consider cases that evaluate to elliptic generalizations, as the status for the conjecture there is currently under discussion \cite{Brown:2020rda}.} (see e.g.\ \cite{Henn:2013pwa,Henn:2014qga}) that any linear combination of Feynman integrals has uniform transcendental weight if it can be written in the form
\begin{equation}
\label{eq:dlogrep}
 \int\sum_i c_i\prod_j\diff\log g_{i,j},
\end{equation}
for $\eps=0$ and with $c_i\in\mathbb{C}$ and $g_{i,j}$ depending on the integration variables, as well as the kinematic variables.
We call $c_i$ the Leading Singularities (LS) of the Feynman integral. Note that eq.\ \eqref{eq:dlogrep} implies the absence of double or higher order poles of the form
\begin{equation}
 \frac{\diff x}{x^a},\qquad a\neq1
\end{equation}
and only permits logarithmic singularities. In the following we will refer to an integrand with this property as a \textit{dlog} integrand, and to the corresponding integral as a \textit{dlog} integral.

In general, the calculation of LS and finding dlog integrals are no easy tasks, but there has been much progress in this direction in recent years \cite{Wasser:2018qvj, Henn:2020lye, Chen:2020uyk}. In \cite{Wasser:2018qvj, Henn:2020lye}, a general algorithm for constructing dlog form integrands in four dimensions has been proposed and the corresponding package has been released publicly. Most importantly, the package also allows to compute the LS of a user-defined integrand and is able to give conditions on an integrand-ansatz for the absence of double poles. We combine this algorithm with the Baikov representation and use an example in section \ref{sec:dbox3m2} to explain in detail how to construct dlog integrals using this approach. This method has already been used in \cite{Chicherin:2018old} to determine the canonical differential equation of a family of five-point integrals.


\subsection{Baikov representation}
\label{sec:dvs4LS}

In both the computation of leading singularities and IBP reduction, the Baikov representation \cite{Baikov:1996cd, Baikov:1996rk, Baikov:2005nv, Lee:2010wea} has proven to be extremely powerful. With $n=LE+L(L+1)/2$ being the number of scalar products, the Baikov representation reads
\begin{equation}
\label{eq:Baikov}
 I_{a_1,\ldots,a_n}=C^L_E U^{\frac{E-D+1}{2}}\int \diff z_1\cdots \diff z_n P^{\frac{D-L-E-1}{2}}\frac{1}{z_1^{a_1}\cdots z_n^{a_n}},
\end{equation}
where the Baikov polynomial is defined as
\begin{equation}
 P=G(l_1,\ldots,l_L,k_1,\ldots,k_E),
\end{equation}
$U$ is a polynomial depending on the external kinematics
\begin{equation}
 U=G(k_1,\ldots,k_E)
\end{equation}
and $C^L_E$ is
\begin{equation}
 C^L_E=J\frac{\pi^{\frac{L-n}{2}}}{\Gamma(\frac{D-E-L+1}{2})\cdots\Gamma(\frac{D-E}{2})},
\end{equation}
with $J$ being a constant Jacobian. The Gram determinant $G$ is defined by
\begin{align}
 G(k_i)&=G(\{k_i\},\{k_i\}),\\
 G(\{k_i\},\{k_j\})&=\det_{i,j}(2k_i\cdot k_j).
\end{align}
The details of the integration contour will not be important for our applications. However, note that by encircling a pole $z_i$ in eq.\ \eqref{eq:Baikov} we see that this representation trivializes the operation of cutting the propagator $z_i\equiv D_i$ of the Feynman integral.

Another way of using the representation \eqref{eq:Baikov} is to apply it to one loop integral at a time, instead of the multi-loop integral as a whole. This is called the loop-by-loop approach \cite{Frellesvig:2017aai} and has the advantage that the number of integration variables is usually lower than in the full Baikov representation. In addition, the integration kernel factorizes into lower-degree polynomials which can be used to construct integrals without double poles \cite{Chen:2020uyk}.

The leading singularity analysis in Baikov representation has another advantage over the conventional analysis in 4D \cite{Chicherin:2018old}: It captures terms in the numerator that vanish when using a four-dimensional parametrization of the loop-momenta. These terms can be constructed from Gram determinants involving more than four internal or external momenta since these naturally vanish in 4D. While one may be tempted to neglect these so-called \textit{D-dimensional upgrades}, the differential equations quickly show that the naive 4D-dlog analysis is not always enough to construct UT integrals. For more details see \cite{Chicherin:2018old}.

\begin{figure}
\centering
\vspace{-1cm}
\includegraphics[scale=0.43]{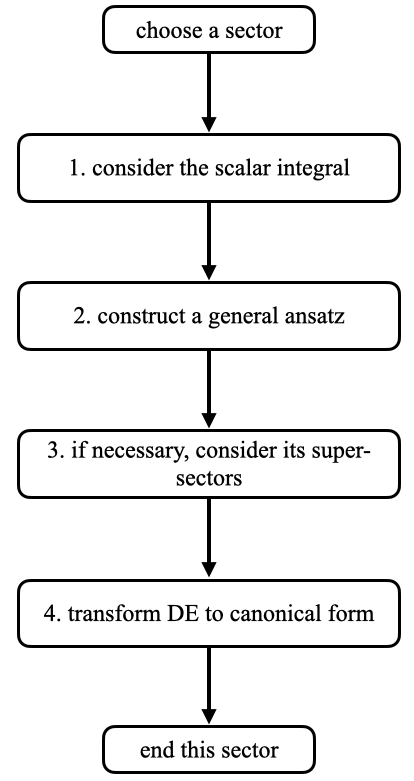}
\caption{A schematic algorithm for finding UTs in a specific sector of a Feynman integral family. In steps one to three, a LS analysis is done.} \label{algorithm}
\end{figure}

The general strategy for finding the UTs is one sector after another in two directions, from top sector to lowest sector or from lowest sector to top sector. The main algorithm for finding dlog integrals or UTs in a specific sector is shown in Fig.\ \ref{algorithm}.
We explain the steps in Fig.\ \ref{algorithm} in more details:
\begin{itemize}
\setlength\itemsep{0.1em}
\item First, we consider the full (standard) or loop-by-loop Baikov representation on the maximal cut of the scalar integral to calculate its LS by using the package \texttt{DlogBasis} from \cite{Wasser:2018qvj, Henn:2020lye}. If there are no double poles, we construct a dlog integral by demanding that the LS are constant.

\item Second, if the scalar integral is not a dlog integral or if the number of MIs in this sector is bigger than one, we consider a general numerator ansatz and repeat the procedure of finding dlog integrals.

\item Third, if the number of dlog integrals we have found is still smaller than the number of MIs, then we consider the super-sectors of the current sector, where we might be able to find more dlog integrals that can be used for the sector.

\item Finally, in general, these dlog integrals are almost UT in the sense that the diagonal blocks of the differential equation, which correspond to the maximal cut, are in canonical form but the off-diagonal blocks might still require corrections. We show how to deal with this case in the next section.
\end{itemize}
The main challenge is the calculation of the LS. This step is especially difficult when more than one square roots appear in the Baikov representation or when the integrand is a function of many kinematic variables or integration variables. The difficulty from square roots is overcome by rationalization through our \texttt{Mathematica} package \cite{RationalizeSquareRoots}.\nocite{Besier:2019kco}

\section{Example: Double box with three massive external legs} 
\label{sec:dbox3m2}

\begin{figure}
\begin{center}
\begin{tikzpicture} [scale=0.7]
\node [left] at (-1.5,-1.5) {$1$};
\draw [ultra thick] (-1.5,-1.5) -- (-1,-1);
\draw [ultra thick, <-] (-1,-1) -- (0,0);
\draw [fill] (0,0) circle [radius=0.05];
\draw [thick, ->] (0,0)--(0,1);
\draw [thick] (0,1)--(0,2);
\draw [fill] (0,2) circle [radius=0.05];
\node [left] at (0,1) {$D_2$};
\draw [ultra thick, ->] (0,2)--(-1,3);
\draw [ultra thick] (-1,3)--(-1.5,3.5);
\node [left] at (-1.5,3.5) {$2$};
\draw [thick,->] (0,2)--(1,2);
\draw [thick] (1,2)--(2,2);
\draw [fill] (2,2) circle [radius=0.05];
\node [above] at (1,2) {$D_3$};
\draw [thick] (2,2)--(3,2);
\draw [thick,<-] (3,2)--(4,2);
\draw [fill] (4,2) circle [radius=0.05];
\node [above] at (3,2) {$D_4$};
\draw [ultra thick, ->] (4,2)--(5,3);
\draw [ultra thick] (5,3)--(5.5,3.5);
\node [right] at (5.5,3.5) {$3$};
\draw [thick] (4,2)--(4,1);
\draw [thick, <-] (4,1)--(4,0);
\draw [fill] (4,0) circle [radius=0.05];
\node [right] at (4,1) {$D_5$};
\draw [thick, ->] (4,0)--(5,-1);
\draw [thick] (5,-1)--(5.5,-1.5);
\node [right] at (5.5,-1.5) {$4$};
\draw [thick] (4,0)--(3,0);
\draw [thick,<-] (3,0)--(2,0);
\node [below] at (3,0) {$D_6$};
\draw [fill] (2,0) circle [radius=0.05];
\draw [thick,->] (2,0)--(1,0);
\draw [thick] (1,0)--(0,0);
\node [below] at (1,0) {$D_1$};
\draw [thick,->] (2,2)--(2,1);
\draw [thick] (2,1)--(2,0);
\node [left] at (2,1) {$D_7$};
\end{tikzpicture}
\end{center}
\caption{The two-loop double box Feynman integral with three massive external momenta $k_1, k_2, k_3$ and seven massless propagators. The massive legs are indicate by thick lines.}
\label{dbox3m}
\end{figure}
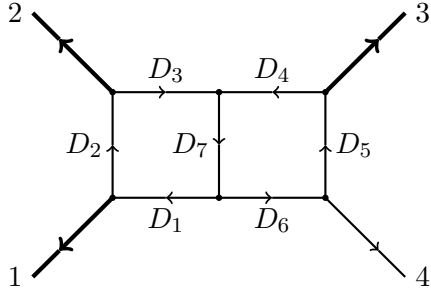

We apply the LS analysis to the double box with three massive external momenta shown in Fig.\ \ref{dbox3m}. We choose this non-trivial example to show the power and procedure of our method. The integral family is defined as
\begin{align}
I_{a_1,\ldots ,a_9}= e^{2 \gamma_{E} \epsilon} \int \frac{d^Dl_1 d^Dl_2}{(i \pi^{d/2})^2} \frac{1}{D_1^{a_1}\cdots D_9^{a_9}},    \label{dbox_3m}
\end{align}
where the propagators are 
\begin{equation}
\begin{aligned}
& D_1=l_1^2,~ D_2=(l_1-k_1)^2,~ D_3=(l_1-k_1-k_2)^2,~ D_4=(l_2+k_1+k_2)^2, \\
& D_5=(l_2+k_1+k_2+k_3)^2,~ D_6=l_2^2,~ D_7=(l_1+l_2)^2,~ \\
& D_8=(l_1-k_1-k_2-k_3)^2,~ D_9=(l_2+k_1)^2,
\end{aligned}
\end{equation}
with the last two being irreducible scalar products (ISPs), i.e.\ $a_8, a_9$ are always non-positive integers. 
The kinematics is 
\begin{align}
k_1^2=m_1^2,~ k_2^2=m_2^2,~ k_3^2=m_3^2~, s=(k_1+k_2)^2,~ t=(k_2+k_3)^2, ~u=(k_1+k_3)^2,
\end{align}
and $s+t+u=\sum_{i=1}^3 m_i^2$. A UT basis for this family has not yet been provided in the literature, but a subset of the integrals have been computed through other methods, see e.g.\ \cite{Usyukina:1993ch}.

After doing IBP reduction with FIRE \cite{Smirnov:2019qkx}, Azurite \cite{Georgoudis:2017iza,Georgoudis:2016wff}, Kira \cite{Klappert:2020nbg} or FiniteFlow \cite{Peraro:2019svx}, we get a total of $47$ master integrals (MIs). The 47 MIs can be organized into $33$ integral sectors as drawn in Fig.\ \ref{allsectors}, and there are respectively $1,3,10,14,5$ sectors with $7,6,5,4,3$ propagators, some of which are symmetric to each other by making permutations of external momenta. There are six slashed box (triangle-triangle) sectors with five propagators shown in the third line of Fig.\ \ref{allsectors}, and among them, one sector is most special with six MIs, while other sectors only contain one or two MIs. 
\input{sectorDrawings.tex}

The task is to find $47$ linearly independent UT integrals, and we will follow the algorithm proposed in Fig.\ \ref{algorithm} in the last section.
We will extensively use the full and the loop-by-loop Baikov representation of eq.\ \eqref{dbox_3m}, therefore we will give the explicit forms of the Baikov representations of our case.
Following eq.\ \eqref{eq:Baikov} we find for the full Baikov representation of the two-loop three-mass box
\begin{align}
\label{dbox3m_full}
I_{a_1,\ldots ,a_9} \sim \int \frac{\diff z_1\cdots \diff z_9}{z_1^{a_1}\cdots z_9^{a_9}} P^{-1-\eps} U^{\eps},
\end{align}
where we have ignored the factor $C^L_E$ because it is not important for our analysis of LS. Here and $P$ and $U$ are
\begin{align}
P=G(l_1,l_2,k_1,k_2,k_3),\qquad U\equiv\Delta=G(k_1,k_2,k_3),
\end{align}
and the scalar products are replaced by new variables corresponding to the propagators $z_i\equiv D_i$. Further, we call 
\begin{align}
\mathcal{K}_{F}(z_1,\ldots,z_9)=P^{-1-\eps} U^{\eps}   \label{full_kernel}
\end{align}
the full Baikov kernel.

Next, we consider the loop-by-loop Baikov representation. In the case of the double box eq.\ \eqref{dbox_3m}, there are two kinds of loop-by-loop Baikov representations \cite{Frellesvig:2017aai}: The left-to-right (LR) and the right-to-left (RL) Baikov representation, depending on whether we first put the $l_1$ or $l_2$ loop into Baikov representation, respectively. The LR Baikov representation is
\begin{align}
\label{dbox3m_LR}
I_{a_1,\ldots, a_7, 0,a_9} \sim \int \frac{\diff z_1\cdots \diff z_7 \diff z_9}{z_1^{a_1}\cdots z_7^{a_7} z_9^{a_9}} \mathcal{K}_{LR}(z_1,\ldots,z_7, z_9)
\end{align}
where the LR Baikov kernel is 
\begin{align}
\mathcal{K}_{LR}(z_1,\ldots,z_7, z_9)=(u_1^{LR})^{-\frac{1}{2}-\eps} (u_2^{LR})^{-\frac{1}{2}-\eps} (u_3^{LR})^{\eps}\Delta^{\eps}
\end{align}
with
\begin{align}
u_1^{LR}=G(l_1,l_2,k_1,k_2),\quad  u_2^{LR}=G(l_2,k_1,k_2,k_3), \qquad u_3^{LR}=G(l_2,k_1,k_2).
\end{align}
The RL Baikov representation is
\begin{align}
\label{dbox3m_RL}
I_{a_1,\ldots, a_8,0} \sim \int \frac{\diff z_1\cdots \diff z_8}{z_1^{a_1}\cdots z_8^{a_8}} \mathcal{K}_{RL}(z_1,\ldots,z_8)
\end{align}
where the RL Baikov kernel is
\begin{align}
\label{eq:RL-kernel}
\mathcal{K}_{RL}(z_1,\ldots,z_8)=(u_1^{RL})^{-\frac{1}{2}-\eps} (u_2^{RL})^{-\frac{1}{2}-\eps} (u_3^{RL})^{\eps} \Delta^{\eps}
\end{align}
and
\begin{align}
\label{eq:RL-grams}
u_1^{RL}=G(l_1,l_2,k_3,k_4),\quad  u_2^{RL}=G(l_1,k_1,k_2,k_3),\quad u_3^{RL}=G(l_1,k_3,k_4).
\end{align}
Just as stated before, the loop-by-loop Baikov representations have fewer integration variables. While this speeds up the LS computation, it also represents a loss of information since integrals involving non-zero $a_8$ or $a_9$, respectively, cannot be included in the ansatz.

\subsection{Top sector and six-propagator sector}

We start with the top sector $\mathcal{I}_{1234567}$, which has $3$ MIs, as shown in Fig.\ \ref{allsectors} 
and we will use the full Baikov representation of eq.\ \eqref{dbox3m_full}.
First, we simply consider the scalar integral $I_{1,1,1,1,1,1,1,0,0}$. After taking the maximal cut, $z_1=\ldots=z_7=0$,
the integrand simplifies to
\begin{equation}
\begin{aligned}
P^{-1-\eps} U^{\eps}|_{z_1=\ldots=z_7=0}& =U^{\epsilon} \Big\{z_8^2 \big[m_1^2m_2^2 s+(-m_1^2s-m_2^2s+s^2)z_9+sz_9^2 \big]  \\
&+ z_8 \big[(m_1^2m_3^2s-s^2t)z_9+(-m_3^2s+s^2)z_9^2 \big] \Big\}^{-1-\epsilon}. \label{top_integrand}
\end{aligned}
\end{equation}
As shown in eq.\ \eqref{eq:dlogrep}
we can set $\eps=0$. The LS is easily calculated by the command \texttt{LeadingSingularities} of the package \cite{Henn:2020lye}:
\begin{align}
-\frac{16}{s(st-m_1^2m_3^2)}.
\end{align}
Therefore the scalar integral $s(m_1^2m_3^2-st) I_{1,1,1,1,1,1,1,0,0}$, whose LS is a rational constant, is a dlog integral on the maximal cut and a candidate for UT in the top sector. 

Next, we need to find two more dlog integrals in the top sector. Since $z_8$ and $z_9$ are the only possible numerators in this sector, we consider the integrals with a numerator $z_8$ and $z_9$, or the linear combinations of them with the scalar integral. Indeed, we find that $I_{1,1,1,1,1,1,1,-1,0}$ and $I_{1,1,1,1,1,1,1,0,-1}$ are dlog integrals on the maximal cut with LS
\begin{align}
-\frac{16}{s(-s+m_3^2)},
\end{align}
and
\begin{align}
-\frac{16}{s r_1},
\end{align}
respectively, where $r_1=\sqrt{\lambda(m_1^2,m_2^2,s)}$ and $\lambda(x,y,z)\equiv x^2+y^2+z^2-2xy-2xz-2yz$ is the K{\"a}ll{\'e}n function. Therefore our UT candidates for the top-sector are
\begin{equation}
\begin{aligned}
& s(m_1^2m_3^2-st) I_{1,1,1,1,1,1,1,0,0},\\
&s(m_3^2-s)I_{1,1,1,1,1,1,1,0,-1},\\
&s r_1I_{1,1,1,1,1,1,1,-1,0} .  \label{topsector}
\end{aligned}
\end{equation}

Some comments: First, the outlined process of finding dlog integrals in the top sector can be simplified, as introduced in \cite{Henn:2020lye}, by giving a general ansatz for the integrand numerator
\begin{align}
(a_1+a_2 z_8+a_3 z_9) \widetilde{\mathcal{K}}_F,
\end{align}
computing all LS, and choosing independent LS to construct UT candidates.
Furthermore, the general ansatz must be linear in $z_8$ and $z_9$, because any higher power would immediately lead to a double pole at infinity. 
Second, although in this sector we take the full Baikov representation as the example to illustrate the process, the loop-by-loop Baikov representation works equally well. Using the LR Baikov representation, we can find the first two dlog integrals of eq. \eqref{topsector}, and the last dlog integral can be found using the RL Baikov representation. When we begin to consider the lower sectors, we will see that the loop-by-loop Baikov representations are usually the better choice, since they contain fewer variables. 

Finally, we need to check if the candidates in eq.\ \eqref{topsector} receive corrections from subsectors when relaxing the maximal cut constraint. We find that for $I_{1,1,1,1,1,1,1,-1,0}$ a second LS appears if $z_1\neq 0$,
\begin{equation}
\label{eq:dlog2LS}
 \frac{16 m_3^2}{s(m_1^2m_3^2+m_2^2s-m_2^2m_3^2-st)}.
\end{equation}
Therefore it is not possible to normalize the integral s.t.\ both LS are simultaneously constant and we need to find suitable subtraction terms to cancel this second Leading Singularity. The subtraction term has to come from the subsector $\mathcal{I}_{234567}$, since this is the only subsector with non-vanishing integrals under the cut $z_2=\ldots=z_7=0$. Making a suitable ansatz,
we find
\begin{equation}
\label{eq:topdlog1}
 r_1(-m_3^2 I_{0,1,1,1,1,1,1,0,0}+sI_{1,1,1,1,1,1,1,-1,0})
\end{equation}
to be a UT integral.
A similar analysis shows that $I_{1,1,1,1,1,1,1,0,-1}$ likewise receives corrections from two subsectors:
\begin{equation}
 (m_3^2-s)(-m_1^2 I_{1,1,1,0,1,1,1,0,0}-m_2^2 I_{1,1,1,1,1,0,1,0,0}+s I_{1,1,1,1,1,1,1,0,-1}).
\end{equation}

Following the same process, i.e.\ considering the maximal cut and calculating the LS, we can easily find the dlog integrals for the sectors with six-propagators:
\begin{equation}
\begin{aligned}
 &(m_1^2m_3^2+m_2^2s-m_2^2m_3^2-st)I_{0,1,1,1,1,1,1,0,0},\\
 &s(m_1^2-t) I_{1,1,0,1,1,1,1,0,0},\\
 &r_2 I_{1,1,1,1,1,0,1,0,0},\label{six_propagator}
\end{aligned}
\end{equation}
where $r_2=\sqrt{(m_2^2m_3^2-m_3^2s+st)^2-4m_1^2m_2^2m_3^2s}$. 
We have checked that lower sectors don't provide any corrections for these UTs. In principle, the maximal cut of the differential equation in these single-MI sectors can always trivially be brought into canonical form by choosing a suitable normalization of the integral. However, this may lead to complicated corrections from lower sectors, therefore we try to avoid it in sectors with many propagators.

\subsection{Five-propagator sectors and lower sectors}
\label{sec:fivep}

At two loops, the sectors with five propagators are usually the most difficult for finding UT integrals. There are three kinds of five-propagator sectors which we call box-bubbles, slashed boxes, and triangle-bubbles as shown in the second, third, and fourth line of Fig.~\ref{allsectors} respectively. First, we consider the six slashed-box sectors. Among them, there are two special sectors, sector $\mathcal{I}_{23567}$ has two MIs and the other sector $\mathcal{I}_{12457}$ has six MIs.

For sector $\mathcal{I}_{23567}$, we immediately find that the scalar integral $I_{0,1,1,0,1,1,1,0,0}$ is a dlog integral with LS
\begin{align}
\frac{16}{m_1^2+m_3^2-s-t}.
\end{align}
 Considering a general ansatz for the numerator in this sector we do not find further dlog integrals. However, as mentioned before, one can often fill missing dlog integrals in a sector with integrals from super-sectors. Again starting from an ansatz in super-sector $\mathcal{I}_{234567}$ under the maximal cut of sector $\mathcal{I}_{23567}$, we find that 
\begin{align}
I_{0,1,1,0,1,1,1,0,0}-I_{0,1,1,1,1,1,1,0,-1}+m_2^2 I_{0,1,1,1,1,1,1,0,0},
\end{align}
has leading singularity
\begin{align}
-\frac{16}{m_3^2-s}
\end{align}
and can therefore be used as a substitute in this sector. Surprisingly, after doing IBP reduction, we find that this integral can be completely reduced to an integral in sector $\mathcal{I}_{23567}$
\begin{align}
I_{0,1,1,0,1,1,1,0,0}-I_{0,1,1,1,1,1,1,0,-1}+m_2^2 I_{0,1,1,1,1,1,1,0,0}= -\frac{m_2^2(t-m_3^2)}{\epsilon (m_3^2-s)} I_{0,2,1,0,1,1,1,0,0}.
\end{align}
Hence we conclude that $\epsilon m_2^2(t-m_3^2) I_{0,2,1,0,1,1,1,0,0}$ is the second UT integral we want. 

Next, we consider sector $\mathcal{I}_{12457}$, which has six MIs and is therefore the sector with the most MIs of the two-loop three-mass box. Although the procedure for this sector is similar to the previous one, it is technically much more difficult. To simplify the computation, we use the loop-by-loop Baikov representations. The schematic form of the maximal cut integrand of the scalar integral $I_{1,1,0,1,1,0,1,0,0}$ in LR Baikov representation is
\begin{align}
\mathcal{K}_{LR}|_{z_1=z_2=z_4=z_5=z_7=0}= \frac{16}{\sqrt{f_1(z_6,z_9)} \sqrt{ f_2(z_3,z_6,z_9)}},  \label{LRkernel_12457}
\end{align}
where we have set $\epsilon=0$. The kernel \eqref{LRkernel_12457} has two complicated square roots and both of their arguments are quadratic polynomials. Fortunately, we can first put the integrand with respect to $z_3$ into dlog form, because it appears only in the second square root. The result is
\begin{align}
  \frac{16}{\sqrt{f_1(z_6,z_9)}\sqrt{f_3(z_6,z_9)}}  \frac{\partial \log g_1(z_3,z_6,z_9) }{\partial z_3}
\end{align}
where $f_3(z_6,z_9)=\lambda(m_1^2,z_6,z_9)$. Since $f_3(z_6,z_9)$ is much simpler now, we can rationalize it through the change of variables\footnote{The command \texttt{LeadingSingularities} cannot find this transformation on its own and therefore stops the computation at this point.}
\begin{align}
z_6 = -m_1^2 x_1(1+x_2), \quad z_9=-m_1^2 x_2(1+x_1),
\end{align}  
which was found through our package \cite{RationalizeSquareRoots}.

The Jacobian turns out to cancel $\sqrt{f_3(z_6(x_1,x_2),z_9(x_1,x_2))}$. 
There is now only one square root left and therefore we can continue the analysis in the usual way to get the LS
\begin{align}
 -\frac{16}{r_4},
\end{align}
with $r_4=\sqrt{(s+t-m_2)^2-4m_1^2m_3^2}=\sqrt{\lambda(m_1^2,m_3^2,u)}$. So the scalar integral $r_4 I_{1,1,0,1,1,0,1,0,0}$ is a dlog integral.

Since we cannot find further dlog integrals through the maximal cut of this sector, we turn to the super-sectors. Through sector $\mathcal{I}_{124567}$ we find that
\begin{align}
(m_3^2-s)(-I_{1,1,0,1,1,0,1,0,0}+I_{1,1,0,1,1,1,1,0,-1}-&m_1^2I_{1,1,0,1,1,1,1,0,0}) \notag \\
=& -\frac{m_1^2(m_3^2-s)}{\epsilon} I_{2,1,0,1,1,0,1,0,0} 
\end{align}
is a UT integral and reduces to sector $\mathcal{I}_{12457}$. To also consider the integrals with $a_8\neq 0$, we switch to the RL Baikov representation and similarly find the UT integral in supersector $\mathcal{I}_{123457}$:
\begin{align}
r_1(I_{1,1,0,1,1,0,1,0,0}&+I_{1,1,1,1,1,0,1,-1,0}-m_3 I_{1,1,1,1,1,0,1,0,0}) 
=-\frac{m_3^2 r_1}{\epsilon} I_{1,1,0,2,1,0,1,0,0}. 
\end{align}
Using the symmetry $p_1\leftrightarrow p_3$ of the sector, we can find another two UT integrals
\begin{align}
\label{eq:hsb-dlog-sym}
 -\frac{m_3^2(m_1^2-t)}{\epsilon} I_{1,1,0,1,2,0,1,0,0}, ~~ -\frac{m_1^2 r_3}{\epsilon} I_{1,2,0,1,1,0,1,0,0}.
\end{align}
From the above results, we guess the last UT integral to be $ I_{1,1,0,1,1,0,2,0,0}$, and its coefficient can be specified by integrating out the $\eps^0$ term of the corresponding differential equation.
Finally, this gives us six UT integrals for this sector as shown in \eqref{allUT}. 
In section \ref{sec:syz} we show an alternative method for finding the first UT integral in eq.\ \eqref{eq:hsb-dlog-sym}.

Now we should consider the remaining sectors, i.e.\ box-bubbles and triangle-bubbles, all of which contain a bubble sub-diagram. Since the bubble integral has a double pole, the scalar integrals in these sectors cannot be dlog integrals. However, we can again consider their super-sectors, where triangle sub-diagram can be reduced to the corresponding bubble. As a result, the bubble integral often comes with a doubled propagator or has an $\eps$ dependent prefactor. We take the sector $\mathcal{I}_{12357}$ as an example. In this sector, we can find that
\begin{align}
-m_3^2I_{1,1,1,1,1,0,1,0,0}+I_{1,1,1,0,1,0,1,-1,0}=-\frac{2\epsilon-1}{\epsilon} I_{1,1,1,0,1,0,1,0,0}
\end{align}
has LS $-16/r_1$, and
\begin{align}
I_{1,1,1,0,1,1,1,0,0}=-\frac{1}{\epsilon}I_{1,1,1,0,1,0,2,0,0}
\end{align}
has LS $16/(m_1^2m_3^2-st)$. Therefore, two UT integrals in this sector are
\begin{align}
\frac{2\epsilon-1}{\epsilon}r_1 I_{1,1,1,0,1,0,1,0,0}\quad \text{and} \quad \frac{1}{\epsilon}(m_1^2m_3^2-st)I_{1,1,1,0,1,0,2,0,0}.
\end{align}
For sectors with four or three propagators, a first UT integral can always immediately be found by considering a doubled propagator in each bubble integral. If necessary, a second UT integral is then found by a transformation of the differential equation. Because there are few subsectors that could give rise to corrections of these integrals, we find this transformation to be extremely simple.

To conclude this section, we have found all UT integrals for the integral family of the double box with three massive legs. They are listed in \eqref{allUT}, and we have checked that their differential equation is indeed in canonical form.
In summary, when starting the LS analysis from the maximally-cut integrals, the most important points are the following:
 \begin{enumerate}
\item The dlog integrals of a sector sometimes receive corrections from its lower sectors, which is indicated by the analysis of LS of the integrals when relaxing the maximal cut of this sector.

\item To find dlog integrals in a given sector, we can consider a general ansatz for the numerator s.t.\ there are no immediate double poles. If we cannot find enough dlog integrals in a specific sector, we continue the search in its super-sectors.

\item Sectors with bubble sub-integrals will always have an inherent double pole. In this case we can continue the analysis with a doubled propagator in the bubble integral or consider integrals in its super-sectors. And if the sector is simple enough, a direct transformation of the differential equation is often the preferred approach.

\item As mentioned at the beginning of the section, the off-diagonal blocks of the differential equation might still require correction terms. By working from lowest sector to highest sector, it is then usually straightforward to also bring these off-diagonal terms into canonical form. This can be done either by integrating out the $\epsilon^0$ term of the off-diagonal block or by going back to point 1.\ and relaxing the cut-conditions for the subsector corresponding to the respective off-diagonal block.

\end{enumerate}

\newpage
The list of all UT integrals, with the ordering corresponding to the classification of sectors, is
\begin{equation}
\begin{aligned}
\mathcal{I}_{1234567}:~
& g_{1}=-\epsilon^2 s(st-m_1^2m_3^2) I_{1,1,1,1,1,1,1,0,0}, \\
& g_{2}= \epsilon^2r_1(-m_3^2 I_{0,1,1,1,1,1,1,0,0}+sI_{1,1,1,1,1,1,1,-1,0}),\\
& g_{3}=\epsilon^2(m_3^2-s)(-m_1^2 I_{1,1,1,0,1,1,1,0,0}-m_2^2 I_{1,1,1,1,1,0,1,0,0}+s I_{1,1,1,1,1,1,1,0,-1}), \\
\mathcal{I}_{234567}:~ 
& g_{4}= \epsilon^2 (m_1^2m_3^2+m_2^2s -m_2^2m_3^2-st)I_{0,1,1,1,1,1,1,0,0}, \\
\mathcal{I}_{124567}:~
& g_{5}=\epsilon^2 s(m_1^2-t) I_{1,1,0,1,1,1,1,0,0}, \\
\mathcal{I}_{123457}:~
& g_{6}=\epsilon^2 r_2 I_{1,1,1,1,1,0,1,0,0}, \\
\mathcal{I}_{24567}:~ 
& g_7=\epsilon(2\epsilon-1)(m_3^2-s) I_{0,1,0,1,1,1,1,0,0}, \\
\mathcal{I}_{12357}:~
& g_8=\epsilon(2\epsilon-1)r_1 I_{1,1,1,0,1,0,1,0,0}, ~~ g_9=\epsilon(m_1^2m_3^2-st) I_{1,1,1,0,2,0,1,0,0}, \\
\mathcal{I}_{12457}:~
&g_{10}=-\epsilon m_1^2 (m_3^2-s) I_{2,1,0,1,1,0,1,0,0}, ~~~g_{11}=-\epsilon m_1^2 r_3 I_{1,2,0,1,1,0,1,0,0}, \\
&g_{12}=-\epsilon m_3^2 r_1 I_{1,1,0,2,1,0,1,0,0}, ~~~g_{13}=-\epsilon m_3^2 (m_1^2-t) I_{1,1,0,1,2,0,1,0,0}, \\
&g_{14}=\epsilon  (m_3^2 m_1^2-st) I_{1,1,0,1,1,0,2,0,0}, ~~~g_{15}= \epsilon^2 r_4 I_{1,1,0,1,1,0,1,0,0}, \\
\mathcal{I}_{23567}:~
& g_{16}= \epsilon^2 (m_1^2+m_3^2-s-t) I_{0,1,1,0,1,1,1,0,0}, ~~~ g_{17}= 2\epsilon m_2^2 (t-m_1^2) I_{0,2,1,0,1,1,1,0,0}, \\
\mathcal{I}_{23467}:~
&g_{18}=\epsilon^2 r_1 I_{0,1,1,1,0,1,1,0,0}, ~~~~\\
\mathcal{I}_{23457}:~
&g_{19}=\epsilon^2 r_3 I_{0,1,1,1,1,0,1,0,0}, \\
\mathcal{I}_{13457}:~
&g_{20}=\epsilon^2(m_3^2-s) I_{1,0,1,1,1,0,1,0,0}, ~~~~\\
\mathcal{I}_{12467}:~
&g_{21}=\epsilon^2 r_1 I_{1,1,0,1,0,1,1,0,0}, \\
\mathcal{I}_{12345}: ~
&g_{22}=\epsilon^2 (m_3^2-s) r_1 I_{1,1,1,1,1,1,0,0,0}, \\
\mathcal{I}_{12346}:  ~
&g_{23}=\epsilon r_1 m_3^2 (-2 \epsilon I_{1,1,1,1,1,1,0,0,0}+s I_{1,1,1,2,1,1,0,0,0}), \\
\mathcal{I}_{2467}:~
& g_{24}=\epsilon r_1 I_{0,2,0,1,0,1,1,0,0}, ~~~
g_{25}= 3 \epsilon(s-m_1^2-m_2^2) I_{0,2,0,1,0,1,1,0,0}+m_2^2 m_1^2 I_{0,2,0,1,0,1,2,0,0}, \\
\mathcal{I}_{2457}:~
& g_{26}= \epsilon r_3 I_{0,2,0,1,1,0,1,0,0}, ~~~
g_{27}= 3 \epsilon(m_3^2-m_2^2-t) I_{0,2,0,1,1,0,1,0,0}+m_2^2 t I_{0,2,0,1,1,0,2,0,0}, \\
\mathcal{I}_{2367}:~
& g_{28}=\epsilon r_1 I_{0,1,1,0,0,2,1,0,0}, ~~~
g_{29}=3 \epsilon(m_2^2-m_1^2-s) I_{0,1,1,0,0,2,1,0,0}+m_1^2 t I_{0,1,1,0,0,2,2,0,0}, \\
\mathcal{I}_{2357}:~
& g_{30}= \epsilon r_3 I_{0,1,1,0,2,0,1,0,0}, ~~~
g_{31}= 3 \epsilon(m_2^2-m_3^2-t) I_{0,1,1,0,2,0,1,0,0}+m_3^2 tI_{0,1,1,0,2,0,2,0,0}, \\
\mathcal{I}_{1247}:~
& g_{32}=\epsilon r_1 I_{1,1,0,2,0,0,1,0,0}, ~~~
 g_{33}=3\epsilon (m_1^2-m_2^2-s)I_{1,1,0,2,0,0,1,0,0}+sm_2^2I_{1,1,0,2,0,0,2,0,0}, \\
\mathcal{I}_{1257}:~
& g_{34}=(2\epsilon-1)(3\epsilon-1) I_{1,1,0,0,1,0,1,0,0}, ~~~\\
\mathcal{I}_{1357}:~
 &g_{35}=(2\epsilon-1)(3\epsilon-1) I_{1,0,1,0,1,0,1,0,0}, \\
\mathcal{I}_{1457}:~
& g_{36}=(2\epsilon-1)(3\epsilon-1) I_{1,0,0,1,1,0,1,0,0}, \\
\mathcal{I}_{2346}:~
& g_{37}=(1-2\epsilon)^2 I_{0,1,1,1,0,1,0,0,0}, ~~~
\mathcal{I}_{2345}:~
 g_{38}=(1-2\epsilon)^2 I_{0,1,1,1,1,0,0,0,0}, \\
\mathcal{I}_{1346}:~
& g_{39}=(1-2\epsilon)^2 I_{1,0,1,1,0,1,0,0,0}, ~~~
\mathcal{I}_{1345}:~
 g_{40}=(1-2\epsilon)^2 I_{1,0,1,1,1,0,0,0,0}, \\
\mathcal{I}_{1246}:~
& g_{41}=(1-2\epsilon)^2 I_{1,1,0,1,0,1,0,0,0}, ~~~
\mathcal{I}_{1245}:
 g_{42}=(1-2\epsilon)^2 I_{1,1,0,1,1,0,0,0,0}, \\
\mathcal{I}_{367}:~ 
& g_{43}=s I_{0,0,2,0,0,2,1,0,0},~~ 
\mathcal{I}_{357}:~ 
 g_{44}=m_3^2 I_{0,0,2,0,2,0,1,0,0},~~~
\mathcal{I}_{267}:~ 
 g_{45}=m_1^2 I_{0,2,0,0,0,2,1,0,0}, \\
\mathcal{I}_{247}:~ 
&g_{46}= t I_{0,2,0,0,2,0,1,0,0}, ~~~
\mathcal{I}_{247}:~ 
g_{47}= m_2^2 I_{0,2,0,2,0,0,1,0,0}.      \label{allUT}
\end{aligned}
\end{equation}
The square roots are defined as
\begin{equation}
\begin{aligned}
 r_1&=\sqrt{\lambda(m_1^2,m_2^2,s)}, ~~r_2=\sqrt{(m_2^2m_3^2-m_3^2s+st)^2-4m_1^2m_2^2m_3^2s}, \\
 r_3&=\sqrt{\lambda(m_2^2,m_3^2,t)}, ~~ r_4=\sqrt{\lambda(m_1^2,m_3^2,u)},
\end{aligned}
\end{equation}
where we have used $s+t+u=\sum_{i=1}^3 m_i^2$. Recall that $\lambda(x,y,z)$ is the K{\"a}ll{\'e}n function.

\subsection{Differential equation and the alphabet}
The alphabet of the three-external-mass double box integral family consists of $30$ symbol letters. Here we explicitly list them:
\begin{itemize}
\item Even letters. There are $15$ even letters:
  \begin{gather}
 W_1=m_1^2,\quad W_2=m_2^2,\quad W_3=m_3^2, \quad W_4=s,\quad W_5=t\quad 
W_6=m_3^2-s,\nonumber\\ 
W_7=m_1^2-t, \quad W_8=m_1^2 m_3^2-s t,\quad W_9=m_1^2+m_3^2-s-t,\nonumber\\W_{10}=m_2^2 s+m_1^2 m_3^2-m_2^2 m_3^2-s t,\nonumber \\W_{11}=-2 m_1^2 s-2 m_2^2 s+m_1^4+m_2^4-2 m_1^2 m_2^2+s^2\nonumber \\W_{12}=-2 m_2^2 t-2 m_3^2 t+m_2^4+m_3^4-2 m_2^2 m_3^2+t^2,\nonumber \\W_{13}=-2 m_2^2 s-2 m_2^2 t+m_2^4-4 m_1^2 m_3^2+s^2+2 s t+t^2,\nonumber \\W_{14}=-2 m_3^2 s^2 t+m_3^4 s^2+2 m_2^2 m_3^2 s t-2 m_2^2 m_3^4 s-4 m_1^2 m_2^2 m_3^2 s+m_2^4 m_3^4+s^2 t^2,\nonumber \\W_{15}=-m_1^2 s t-m_2^2 s t-m_3^2 s t+m_1^2 m_2^2 s-m_1^2 m_3^2 s-m_1^2 m_3^2 t+m_2^2 m_3^2 t\nonumber \\+m_1^2 m_3^4+m_1^4 m_3^2-m_1^2 m_2^2 m_3^2+s^2 t+s t^2
  \end{gather}
\item Odd letters. There are $15$ odd letters.
  \begin{gather}
    W_{16}=\frac{m_1^2-m_2^2+s-r_1}{m_1^2-m_2^2+s+r_1}, \quad W_{17}=\frac{m_2^2-m_3^2+t-r_3}{m_2^2-m_3^2+t+r_3},\nonumber\\
W_{18}=\frac{m_1^2-m_2^2-s-r_1}{m_1^2-m_2^2-s+r_1}, \quad
W_{19}=\frac{m_2^2-m_3^2-t-r_3}{m_2^2-m_3^2-t+r_3},\nonumber\\
W_{20}=\frac{m_2^2-s-t+r_4}{m_2^2-s-t-r_4},\quad 
W_{21}=\frac{-m_3^2 s+m_2^2 m_3^2+s t+r_2}{-m_3^2 s+m_2^2 m_3^2+s t-r_2},\nonumber \\
W_{22}=\frac{-r_4 \left(m_3^2-s\right)+m_2^2 s+m_3^2 s-m_3^2 t+2 m_1^2 m_3^2-m_2^2 m_3^2-s^2-s t}{r_4 \left(m_3^2-s\right)+m_2^2 s+m_3^2 s-m_3^2 t+2 m_1^2 m_3^2-m_2^2 m_3^2-s^2-s t},\nonumber\\
W_{23}=\frac{m_2^2 s-m_3^2 s+m_2^2 t-m_1^2 t-m_2^4+m_1^2 m_2^2+m_3^2 m_2^2+m_1^2 m_3^2+s t-r_1 r_3}{m_2^2 s-m_3^2 s+m_2^2 t-m_1^2 t-m_2^4+m_1^2 m_2^2+m_3^2 m_2^2+m_1^2 m_3^2+s t+r_1 r_3},\nonumber\\
W_{24}=\frac{-m_2^2 s+m_3^2 s-2 m_2^2 t-m_3^2 t+m_2^4-2 m_1^2 m_3^2-m_2^2 m_3^2+s t+t^2-r_3 r_4}{-m_2^2 s+m_3^2 s-2 m_2^2 t-m_3^2 t+m_2^4-2 m_1^2 m_3^2-m_2^2 m_3^2+s t+t^2+r_3 r_4},\nonumber\\
W_{25}=\frac{m_1^2 s+2 m_2^2 s-m_1^2 t+m_2^2 t-m_2^4+m_1^2 m_2^2+2 m_1^2 m_3^2-s^2-s t-r_1 r_4}{m_1^2 s+2 m_2^2 s-m_1^2 t+m_2^2 t-m_2^4+m_1^2 m_2^2+2 m_1^2 m_3^2-s^2-s t+r_1 r_4},\nonumber
\end{gather}
\begin{gather}
W_{26}=\frac{f_{26}-r_1r_2}{f_{26}+r_1r_2},\quad W_{27}=\frac{f_{27}-r_2r_3}{f_{27}+r_2r_3},\quad W_{28}=\frac{f_{28}-(m_1^2 m_3^2 - s t)r_2}{f_{28}+(m_1^2 m_3^2 - s t)r_2},\nonumber\\
W_{29}=\frac{f_{29}-(m_2^2 s+m_1^2 m_3^2-m_2^2 m_3^2-s t)r_1}{f_{29}+(m_2^2 s+m_1^2 m_3^2-m_2^2 m_3^2-s t)r_1},\nonumber\\
W_{30}=\frac{f_{30}-(-m_2^2 t+m_1^2 m_2^2-m_1^2 m_3^2+s t)r_3}{f_{30}+(-m_2^2 t+m_1^2 m_2^2-m_1^2 m_3^2+s t)r_3}
  \end{gather}
where 
\begin{gather}
  f_{26}=-m_3^2 s^2-m_1^2 s t-m_2^2 s t+2 m_1^2 m_2^2 s+m_1^2 m_3^2 s\nonumber\\+2 m_2^2 m_3^2 s-m_2^4 m_3^2+m_1^2 m_2^2 m_3^2+s^2 t,\nonumber\\
f_{27}=m_2^2 s t+2 m_3^2 s t-m_3^4 s+m_2^2 m_3^2 s-m_2^2 m_3^2 t\nonumber\\+m_2^2 m_3^4-m_2^4 m_3^2+2 m_1^2 m_2^2 m_3^2-s t^2,\nonumber\\
f_{28}=-m_3^2 s^2 t-m_1^2 m_3^2 s t+m_2^2 m_3^2 s t+m_1^2 m_3^4 s\nonumber\\-2 m_1^2 m_2^2 m_3^2 s+m_1^2 m_2^2 m_3^4+s^2 t^2,\nonumber\\
f_{29}=m_2^2 s^2-m_1^2 s t+m_2^2 s t-m_2^4 s+m_1^2 m_2^2 s-m_1^2 m_3^2 s\nonumber\\-m_2^2 m_3^2 s+m_1^4 m_3^2+m_2^4 m_3^2-2 m_1^2 m_2^2 m_3^2+s^2 t,\nonumber\\
f_{30}=m_2^2 s t-m_3^2 s t+m_2^2 t^2-m_2^4 t-m_1^2 m_2^2 t-m_1^2 m_3^2 t\nonumber\\+m_2^2 m_3^2 t+m_1^2 m_2^4+m_1^2 m_3^4-2 m_1^2 m_2^2 m_3^2+s t^2
\end{gather}
\end{itemize}

In terms of the $30$ letters, the differential equation can be expressed as
\begin{gather}
  d I= \epsilon (d\tilde A)I
\end{gather}
where each entry of the matrix $\tilde A$ is a linear combination of the logarithm of the letters. The computer-readable files for the canonical differential equation and the alphabet are given as the auxiliary files of this paper. 

We also explicitly checked that the {\it one-loop} three-external-mass box's alphabet is a subset of our $30$-letter alphabet.


Obtaining the matrix $\tilde{A}$ from the five matrices $\partial_i\tilde{A}$ of partial differential equations is complicated by the fact that we were not able to find a parametrization that simultaneously rationalizes the four square roots $r_1,\ldots,r_4$. However, individual elements of $\tilde{A}$ only depend on two of the square roots at most and therefore we can use different parametrizations when integrating different elements. E.g.\ $\tilde{A}_{44,21}$ depends only on $r_1$ and $r_2$ and can be integrated after rationalizing only these two square roots using \cite{RationalizeSquareRoots}. In this way we are able to rationalize all elements of the matrices $\partial_i\tilde{A}$. Subsequently, we obtain the independent letters following Sec.\ 3.1 of \cite{Bonciani:2019jyb}.

\section{Reducing shifted integrals to Feynman integrals through syzygies}
\label{sec:syz}



In this section we try to utilize a different idea for finding UT integrals which was introduced in \cite{Chen:2020uyk}, and we further develop this method by a new type of IBP relations in the loop-by-loop Baikov representation. The method of \cite{Chen:2020uyk} builds on the fact that the Baikov kernel $\mathcal{K}(\bz)$ in the loop-by-loop approach factorizes in the following way:
\begin{equation}
\label{eq:u-factorization}
 \mathcal{K}(\bz)=\prod_{k=1}^m u_k(\bz_k)^{-\gamma_k-\beta_k\eps},
\end{equation}
where it is important that the $u_k(\bz_k)$ are irreducible, distinct and quadratic polynomials in $\bz=\{z_1,\ldots,z_n\}$. For the example of the two-loop three-mass box in the right-to-left approach, the kernel is given in eq.\ \eqref{eq:RL-kernel}
\begin{equation}
\label{eq:K-kernel}
 \mathcal{K}(\bz)=\left(\frac{\Delta\, u_3(\bz_3)}{u_1(\bz_1)u_2(\bz_2)}\right)^{\eps} \frac{1}{\sqrt{u_1(\bz_1)}\sqrt{u_2(\bz_2)}}
\end{equation}
where $\bz=\{z_1,z_2,z_3,z_4,z_5,z_6,z_7,z_8\}$ and $\bz_1=\bz\setminus \{z_2\},\bz_2=\{z_1,z_2,z_3,z_8\},\bz_3=\{z_1,z_3,z_8\}$.
For the sake of readability we use $u_i(\bz_i)\equiv u_i^{RL}$ and in the following also $u_i\equiv u_i(\bz_i)$. From eq.\ \eqref{eq:K-kernel} we see that
\begin{equation}
 \gamma_1=\frac{1}{2},\qquad \gamma_2=\frac{1}{2},\qquad \gamma_3=0.
\end{equation}
In the last section we used the square-roots of $u_1(\bz_1)$ and $u_2(\bz_2)$ in the denominator to build dlog-integrands by choosing appropriate numerators.
Following \cite{Chen:2020uyk} we now also want to introduce $u_3(\bz_3)$ into the denominator and likewise use it in the dlog construction. From eq.\ \eqref{eq:u-factorization} we see that this amounts to the shift $\gamma_3\rightarrow(\gamma_3+1)$\footnote{Note that we can only relate integrals with $\gamma_k$'s shifted by an integer back to the unshifted integrals. Therefore, we can only shift $\gamma_3$ without introducing a double-pole into the integrand.}. We will use the notation $I^{(\gamma_1,\gamma_2,\gamma_3+1)}$ for these new integrals, s.t.\ the original integrals are denoted by $I^{(\gamma_1,\gamma_2,\gamma_3)}\equiv I$.

The task is then to
\begin{enumerate}
 \item Find dlog integrands through suitable linear combinations of integrals $I^{(\gamma_1,\gamma_2,\gamma_3+1)}$ and $I^{(\gamma_1,\gamma_2,\gamma_3)}$.
 \item Relate the shifted integrals $I^{(\gamma_1,\gamma_2,\gamma_3+1)}$ to the unshifted integrals $I^{(\gamma_1,\gamma_2,\gamma_3)}$.
\end{enumerate}
In both of these points, we will deviate from the original methods proposed in \cite{Chen:2020uyk}. In the first point, the two square roots in $\mathcal{K}(\bz)$ force us to apply a change of variables that rationalizes one of the them. In the second point, we will not use intersection theory \cite{Mizera:2017rqa,Mastrolia:2018uzb,Frellesvig:2019kgj,Frellesvig:2019uqt,Weinzierl:2020xyy}, but rely on unconventional IBP reduction via the Laporta algorithm, in combination with techniques borrowed from dimensional recurrence relations \cite{Lee:2009dh}.

\subsection{Finding dlog integrands}
\label{sec:dlogssyz}

It will be useful to define
\begin{equation}
 u_{i,j_1^\pm\ldots j_m^\pm}=u_i(z_{j_1}= c_{j_1}^\pm,\ldots,z_{j_m}= c_{j_m}^\pm),
\end{equation}
with $c_j^\pm=0, j\notin\{3,6,8\}$ and $c_3^\pm,c_6^\pm$ and $c_8^\pm$ will be defined below depending on the sector under consideration.

Next, we note that, when constructing dlog integrals, a factor $(x-a)$ in front of a square root with a quadratic polynomial as argument basically means that we take the residue in $a$, because \cite{Chen:2020uyk}
\begin{align}
\label{eq:sqrt1}
 \frac{1}{(x-a)\sqrt{(x-a_1)(x-a_2)}}&=\frac{1}{\sqrt{(a-a_1)(a-a_2)}}\ \frac{\partial}{\partial x}\log\frac{1+\sqrt{\frac{(a_2-a)(a_1-x)}{(a_1-a)(a_2-x)}}}{1-\sqrt{\frac{(a_2-a)(a_1-x)}{(a_1-a)(a_2-x)}}}\\
 \frac{1}{(x-a)\sqrt{(x-x_1)}}&=\frac{1}{\sqrt{(a-x_1)}}\ \frac{\partial}{\partial x}\log\frac{1+\sqrt{\frac{(a_1-x)}{(a_1-a)}}}{1-\sqrt{\frac{(a_1-x)}{(a_1-a)}}}
\end{align}
and we usually omit the dlog because its exact form is irrelevant for the construction.
This means that, for the top sector, one can immediately construct dlogs in $z_4,z_5,z_6$ and $z_7$ by repeatedly using eq.\ \eqref{eq:sqrt1}, effectively setting these variables to zero in $u_1$. The result is
\begin{equation}
 I_{1,1,1,1,1,1,1,0,0}\sim\frac{1}{z_1z_2z_3z_4z_5z_6z_7}\frac{1}{\sqrt{u_1} \sqrt{u_2}}=\frac{1}{z_1z_2z_3}\frac{1}{\sqrt{u_{1,4567}}\sqrt{u_2}}\times\prod_{i=4}^7\frac{\partial}{\partial z_i}\log(\ldots)
\end{equation}
where
\begin{equation}
u_{1,4567}=(m_3^2 z_1-s z_8)^2
\end{equation}
which rationalizes the first square root. This is possible because $u_2$ does not depend on these four variables.

A more interesting example is the ``hard slashed-box'' sector, $\mathcal{I}_{12457}$. As a warm-up, consider again the scalar integral, this time in the RL approach
\begin{equation}
 I_{1,1,0,1,1,0,1,0,0}\sim\frac{1}{z_1z_2z_4z_5z_7}\frac{1}{\sqrt{u_1}\sqrt{u_2}}.
\end{equation}
We can set $z_4, z_5$ and $z_7$ to zero in $u_1$, but because we cannot set $z_6$ to zero, the square-root does not rationalize anymore. Next, we can factorize the remaining $u_{1,457}$ in $z_6$:
\begin{equation}
\label{eq:G1scalar}
 u_{1,457}=f_1(z_3,z_8) (z_6-c_6^+)(z_6-c_6^-)
\end{equation}
and build a dlog in $z_6$ by using
\begin{equation}
\label{eq:sqrt2}
 \frac{1}{\sqrt{(x-a_1)(x-a_2)}}=\frac{\partial}{\partial x}\log\frac{1+\sqrt{\frac{(a_1-x)}{(a_2-x)}}}{1-\sqrt{\frac{(a_1-x)}{(a_2-x)}}},
\end{equation}
so that only $z_1z_2\sqrt{f_1(z_3,z_8)}\sqrt{u_2}$ remains in the denominator.

Similarly, we can set $z_1$ and $z_2$ in $u_2$ to zero, but not $z_3$. Hence, for the scalar integral, we are left with
\begin{equation}
\label{eq:hsb-scalardlog}
 \frac{1}{\sqrt{f_1(z_3,z_8)}\sqrt{u_{2,12}}}\times\prod_{i\in\{1,2,4,5,6,7\}}\frac{\partial}{\partial z_i}\log(\ldots).
\end{equation}
Since both factors are quadratic in $z_3$ and $z_8$, we need to rationalize one of the square roots. We can rationalize $f_1(z_3,z_8)$ through the change of variables
\begin{equation}
\label{eq:rativars}
 z_3=-m_3^2x_3(1+x_8),\qquad z_8=-m_3^2(1+x_3)x_8,
\end{equation}
which can be found through the methods described in \cite{Besier:2018jen}.
The Jacobian turns out to cancel $\sqrt{f_1(x_3,x_8)}$ and therefore there is only $\sqrt{u_{2,12}(x_3,x_8)}$ left, which is quadratic in $x_3$ and $x_8$ and can be processed through eq.\ \eqref{eq:sqrt2}. This shows how the scalar integral in this sector can be put into dlog form using the language of \cite{Chen:2020uyk}. It also shows why this integral does not receive corrections from lower sector, namely because the analysis on the maximal cut in eq.\ \eqref{LRkernel_12457} is essentially equivalent to the one done in this section.

Now we are ready to search for further dlog integrals among the shifted integrals $I^{(\gamma_1,\gamma_2,\gamma_3+1)}$. We consider the integrand of
\begin{equation}
 I^{(\gamma_1,\gamma_2,\gamma_3+1)}_{1,1,0,1,1,0,1,0,0}\sim\frac{1}{z_1z_2z_4z_5z_7}\frac{1}{\sqrt{u_1}\sqrt{u_2}\, u_3}.
\end{equation}
The steps for $u_1$ are the same up to eq.\ \eqref{eq:G1scalar}, i.e.\ we begin by building dlogs in $z_4,z_5,z_7$ and then in $z_6$. For $u_2$, we do not yet set $z_1$ and $z_2$ to zero but first change variables according to eq.\ \eqref{eq:rativars} which removes $\sqrt{f_1(x_3,x_8)}$. The integrand up to now is therefore
\begin{equation}
 I^{(\gamma_1,\gamma_2,\gamma_3+1)}_{1,1,0,1,1,0,1,0,0}\sim\frac{1}{z_1z_2}\frac{1}{\sqrt{u_2}\, u_3}\times\prod_{i\in\{4,5,6,7\}}\frac{\partial}{\partial z_i}\log(\ldots).
\end{equation}

Now we need to factorize $u_3$, e.g.\
\begin{equation}
\label{eq:u3fact}
u_3(z_1,x_3,x_8)=f_3(x_8)(x_3-c_3^+)(x_3-c_3^-). 
\end{equation}
To make the discussion easier, it will be beneficial to remove one of the two factors in $x_3$ by switching to an integral $I^{(\gamma_1,\gamma_2,\gamma_3+1)}_{1,1,0,1,1,0,1,0,0}\left[\mathcal{N}_{\pm}\right]$ with corresponding numerator $\mathcal{N}_+=(x_3-c_3^-)$ or $\mathcal{N}_-=(x_3-c_3^+)$. The complete denominator in \eqref{eq:u3fact} can then be recovered by building linear combinations of the two integrals with numerators. Therefore, we consider the integrand
\begin{equation}
\begin{aligned}
\label{eq:intgrnd3}
 I^{(\gamma_1,\gamma_2,\gamma_3+1)}_{1,1,0,1,1,0,1,0,0}\left[\mathcal{N}_{\pm}\right]&=\frac{1}{z_1z_2f_3(x_8)(x_3-c_3^\pm)\sqrt{u_{2}}} \\
 &=\frac{1}{z_1z_2f_3(x_8)\sqrt{u_{2,3^\pm}}}\times\frac{\partial}{\partial x_3}\log(\ldots)\times\prod_{i\in\{4,5,6,7\}}\frac{\partial}{\partial z_i}\log(\ldots).
 \end{aligned}
\end{equation}

As shown in appendix \ref{sec:Gramrels}, $u_{2,3^\pm}$ is a perfect square and can be written in the following way:
\begin{equation}
\label{eq:gramder}
 \sqrt{u_{2,3^\pm}}=\frac{1}{(m_3^2-s)}\frac{1}{2}\left[1-(x_3-c_3^\pm)\frac{\partial}{\partial x_3}\right]\frac{\partial}{\partial z_2}u_2.
\end{equation}
We could now use $\sqrt{u_{2,3^\pm}}$ in the numerator in order to cancel the corresponding denominator of \eqref{eq:intgrnd3}. This would leave us with dlogs in $z_1,z_2$ and $x_8$, as one can easily check. However, $c_3^\pm$ are potential square roots and the relation between $\{x_3,x_8\}$ and $\{z_3,z_8\}$ is not single-valued. The numerators $\mathcal{N}^{\prime}_{\pm}=\sqrt{u_{2,3^\pm}}(x_3-c_3^\mp)$ are therefore also not single-valued. To avoid this issue, we build linear combinations of the two numerators.

The anti-symmetric combination is
\begin{equation}
\begin{aligned}
\label{eq:antisymG3}
 \mathcal{N}^\prime_+-\mathcal{N}^\prime_-&=\sqrt{u_{2,3^+}}(x_3-c_3^-)-\sqrt{u_{2,3^-}}(x_3-c_3^+)\\
 &=\frac{1}{2(m_3^2-s)}(c_3^+-c_3^-)\frac{\partial}{\partial z_2}u_2.
 \end{aligned}
\end{equation}
The factor $c_3^+-c_3^-$ can easily be dealt with by cancelling it through a corresponding denominator. This means that the two individual integrals of the linear combination now have numerators $\mathcal{N}^{\prime\prime}_\pm=\sqrt{u_{2,3^\pm}}(x_3-c_3^\mp)/(c_3^+-c_3^-)$ and $(c_3^+-c_3^-)$ is used in the dlog construction for $z_1$. The final integrand of the anti-symmetric combination is then
\begin{equation}
\label{eq:antisymcomint}
 I^{(\gamma_1,\gamma_2,\gamma_3+1)}_{1,1,0,1,1,0,1,0,0}[\mathcal{N}]\sim\frac{1}{z_1z_2z_4z_5z_7}\frac{\mathcal{N}}{\sqrt{u_1}\sqrt{u_2}\, u_3},
\end{equation}
where the numerator is now a rational function in $\bz$
\begin{equation}
 \mathcal{N}=2\left(\mathcal{N}^{\prime\prime}_+-\mathcal{N}^{\prime\prime}_-\right)(m_3^2-s)^2=(m_3^2-s)\frac{\partial}{\partial z_2}u_2.
\end{equation}

Making the symmetric combination rational in $z_3$ and $z_8$ is more complicated. While this can be done through certain linear combinations, e.g.\ using
\begin{equation}
 \frac{\partial}{\partial x_3}-\frac{\partial}{\partial x_8}=m_3^2\left(\frac{\partial}{\partial z_3}-\frac{\partial}{\partial z_8}\right),
\end{equation}
we find it both easier and more systematic to just make an ansatz in terms of integrals $I^{(\gamma_1,\gamma_2,\gamma_3)}$ and $I^{(\gamma_1,\gamma_2,\gamma_3+1)}$ and then feed it into \texttt{LeadingSingularities}. In this way, we find a total of six dlog integrals, including the two mentioned above.

\subsection{Reducing shifted integrals}

Our method for relating the shifted integrals $I^{(\gamma_1,\gamma_2,\gamma_3+1)}$ to the original ones is similar to how dimensional recurrence relations are derived. Therefore, we give a very brief introduction into the latter before proceeding with the example of the last section.

Following eq.\ (8) in \cite{Lee:2009dh}, one can construct lowering dimensional recurrence relations in (full) Baikov representation by factoring out one power of the Baikov polynomial:
\begin{align}
  I^{(D)}_{a_1,\ldots,a_n}&\sim\int\left(\prod_{i=1}^{n}\diff z_i\right)\frac{P(z_1,\ldots,z_n)^{(D-E-L-1)/2}}{z_1^{a_1}\ldots z_n^{a_n}}\\
  \label{eq:lowerD}
  I^{(D+2)}_{a_1,\ldots,a_n}&\sim P(A^-_1,\ldots,A^-_n)\,I^{(D)}_{a_1,\ldots,a_n},
\end{align}
where $A^-_i$ acts on $I^{(D)}_{a_1,\ldots,a_n}$ by decreasing the corresponding $a_i$ by one, e.g.\
\begin{equation}
 A_3^-\, I^{(D)}_{a_1,a_2,a_3,a_4}=I^{(D)}_{a_1,a_2,a_3-1,a_4}.
\end{equation}
An important point here is that one can do a $(D+2)$-dimensional IBP reduction for $I^{(D+2)}_{a_1,\ldots,a_n}$ and derive the corresponding master integrals $\bm{I}^{(D+2)}_{\text{MI}}$, which can then be translated to integrals in $D$ dimensions. These integrals can then be related to $D$-dimensional master integrals $\bm{I}^{(D)}_{\text{MI}}$ through a $D$-dimensional IBP reduction. This gives a relation
\begin{equation}
\label{eq:lowerDMI}
 \bm{I}^{(D+2)}_{\text{MI}}=T_{-2}\,\bm{I}^{(D)}_{\text{MI}}.
\end{equation}
Upon inverting $T_{-2}$ one can also derive raising dimensional recurrence relations, which have to be consistent with the ones found from the parametric representation (see eq.\ (4) in \cite{Lee:2009dh}).

Now the idea is to apply this to the right-to-left Baikov approach for a single $u_j(\bz_j)$. We start with an analog expression for the integral, now in the loop-by-loop approach:
\begin{align}
\label{eq:I-family}
 I^{(\gamma_1,\ldots,\gamma_m)}_{a_1,\ldots,a_n}&\sim\int\left(\prod_{i=1}^{n}\diff z_i\right)\frac{\prod_{k=1}^m u_k(\bz_k)^{-\gamma_k-\beta_k\eps}}{z_1^{a_1}\ldots z_n^{a_n}}
\end{align}
As in section \ref{sec:dlogssyz}, our dlog integrals will have an additional factor $u_j(\bz_j)$ in the denominator, i.e.\ $\gamma_j\rightarrow(\gamma_j+1)$. We relate the two types of integrals through.
\begin{equation}
 I^{(\gamma_1,\ldots,\gamma_m)}_{a_1,\ldots,a_n}= u_j(A^-_1,\ldots,A^-_n)\,I^{(\gamma_1,\ldots,\gamma_j+1,\ldots,\gamma_m)}_{a_1,\ldots,a_n},
\end{equation}
We would now again like to do an IBP reduction of both sides to get
\begin{equation}
 \bm{I}^{(\gamma_1,\ldots,\gamma_m)}_{\text{MI}}=T_{\gamma_j+1}\,\bm{I}^{(\gamma_1,\ldots,\gamma_j+1,\ldots,\gamma_m)}_{\text{MI}}
\end{equation}
and use this to relate the integrals $I^{(\gamma_1,\ldots,\gamma_j+1,\ldots,\gamma_m)}_{a_1,\ldots,a_n}$ to the master integrals $\bm{I}^{(\gamma_1,\ldots,\gamma_m)}_{\text{MI}}$.

However, in general it is not expected that an integral with denominators other than $z_i^{a_i}$ can be related to the integral family \eqref{eq:I-family}. Indeed, we find that $T_{\gamma_j+1}$ is not invertible and that only a subset of the $\bm{I}^{(\gamma_1,\ldots,\gamma_j+1,\ldots,\gamma_m)}_{\text{MI}}$ can be written in terms of $\bm{I}^{(\gamma_1,\ldots,\gamma_j,\ldots,\gamma_m)}_{\text{MI}}$. Applying this to our example, we find only two linear combinations of the dlog integrals of the previous section for which this is possible.
The first integral is simply the scalar integral, and the second integral is
\begin{equation}
\begin{aligned}
 (m_1^2-t)&\big[-I_{1,1,0,1,1,0,1,0,0}+(m_3^2-s)I^{(\gamma_1,\gamma_2,\gamma_3+1)}_{1,1,-2,1,1,0,1,0,0}+s I^{(\gamma_1,\gamma_2,\gamma_3+1)}_{1,1,-1,1,1,0,1,-1,0}\\
 &-(m_3^2-s)(m_3^2+s)I^{(\gamma_1,\gamma_2,\gamma_3+1)}_{1,1,-1,1,1,0,1,0,0}-m_3^2 s I^{(\gamma_1,\gamma_2,\gamma_3+1)}_{1,1,0,1,1,0,1,-1,0}\\
 &+m_3^2(m_3^2-s)s I^{(\gamma_1,\gamma_2,\gamma_3+1)}_{1,1,0,1,1,0,1,0,0}\big]\\
 &=-\eps^{-1}m_3^2(m_1^2-t)I_{1,1,0,1,2,0,1,0,0}.
 \end{aligned}
\end{equation}

At this point, we want to comment on
the possibility of writing integrals with additional denominators $u_i$ in terms of the original integral family.
Looking at eq.\ \eqref{eq:antisymcomint} we see that our construction would favor $u_2$, rather than $u_3$, in the denominator because the derivative in $z_2$ would combine into $\partial_{z_2}\mathrm{log}\,u_2$. This, in turn, would make the removal of the denominator trivial, because of IBPs of the form
\begin{equation}
 0=\diff\left[\frac{\diff z_1\ldots\widehat{\diff z_2}\ldots\diff z_8\, \mathcal{K}(\bz)}{z_1z_2z_4z_5z_7} \right]=(\gamma_2+\eps\beta_2)\frac{\diff^8\bz\ \mathcal{K}(\bz)}{z_1z_2z_4z_5z_7}\partial_{z_2}\log\,u_2+\frac{\diff^8\bz\ \mathcal{K}(\bz)}{z_1z_2^2z_4z_5z_7}.
\end{equation}
This happens e.g.\ in the two-loop four-scale triangle integrals discussed in \cite{Chen:2020uyk}.

\subsection{Details about the IBP reduction in the loop-by-loop approach}

A general IBP in the loop-by-loop approach has the form
\begin{equation}
 \begin{aligned}\label{eq:basicIBP}
  0&=\diff\left(\sum_{i=1}^n (-1)^{i+1}v_i\, \mathcal{K}(\bz)\frac{\diff z_1\ldots\widehat{\diff z_i}\ldots\diff z_n}{z_1^{a_1}\ldots z_n^{a_n}} \right)\\
  &=\left[\sum_{i=1}^n\left(\partial_{z_i}v_i+v_i\,\partial_{z_i}\log \mathcal{K}(\bz)-a_i\frac{v_i}{z_i}\right)\right]\Omega
 \end{aligned}
\end{equation}
where $\Omega=\mathcal{K}(\bz)\diff z_1\ldots\diff z_n/(z_1^{a_1}\ldots z_n^{a_n})$. The derivative of $\mathcal{K}(\bz)$ is
\begin{equation}
 \partial_{z_i}\log \mathcal{K}(\bz)=\sum_{k=1}^m(-\gamma_k-\beta_k\eps)\frac{\partial_{z_i}u_k(\bz_k)}{u_k(\bz_k)},
\end{equation}
where we made use of the factorization into irreducible factors in eq.\ \eqref{eq:u-factorization}.
Therefore
\begin{equation}
 \sum_{i=1}^n v_i\partial_{z_i}\log \mathcal{K}(\bz)=\sum_{k=1}^m\frac{(-\gamma_k-\beta_k\eps)}{u_k(\bz_k)}\sum_{i=1}^n v_i\partial_{z_i}u_k(\bz_k).
\end{equation}
To get IBPs without shifts in the $\gamma_k$, we demand that the corresponding numerators are proportional to the denominators:
\begin{equation}
 b_ku_k+\sum_{i=1}^n v_i\partial_{z_i}u_k(\bz_k)=0,\qquad \forall k
\end{equation}
which amounts to computing the syzygy of the module consisting of the $u_k(\bz_k)$ and their derivatives. The syzygy of the module can be easily solved by the computational algebraic geometry software Singular \cite{DGPS}. The resulting IBP vectors $v_i$ can then be used in eq.\ \eqref{eq:basicIBP} to build a system of IBP relations without positive shifts in the $\gamma_k$.

\section{Summary and outlook} 
\label{sec:summary}
In this paper, we further develop the UT integral determination method based on the leading singularity analysis of Feynman integrals in Baikov representation. For an integral family, we analyze sector by sector, rationalize the integrand factors in Baikov representation if necessary, and then use computational tools for leading singularities to find UT integral candidates. After the sector-by-sector analysis, we collect the UT integral candidates, use IBP identities to determine the independent integrals and derive the differential equation to verify the UT candidates.

Our method is powered by the state-of-art tools for the integrand leading singularity and dlog form computations, and for the rationalization of radicals. This method is highly flexible and easy to use since 1) for each sector we can use either the original Baikov or loop-by-loop Baikov representation 2) we may trade ``complicated'' UT integrals in some sectors with reducible UT integrals in the corresponding super sectors, in the sector-by-sector leading singularity analysis, to make the UT searching easier.  

We present a nontrivial example, the UT integral basis for the two-loop double box family with three different external mass, to demonstrate the power of our method. The complete UT basis is determined by our Baikov leading singularity analysis combined with our package for rationalization. The corresponding canonical differential equation is obtained and the alphabet is determined by integrating the differential equation with the help of our rationalization package.

Furthermore, we developed the IBP reduction method for the loop-by-loop Baikov representation. In general, the loop-by-loop Baikov representation contains several irreducible factors and thus we can use the syzygy module method to find the IBP relations in this type of integrand. What is more, since it is clear that there are dlog integrals in this representation which are {\it not} obviously of the form of Feynman integrals, we can fine tune our syzygy equations to convert these integrals to Feynman integrals and then get UT candidates.

In the future, we plan to make a package with our UT determination protocol based on the leading singularity analysis in Baikov representation, to automize the steps of rationalization, leading singularity analysis, and IBP reduction for differential equations.

\acknowledgments

We thank Alessandro Georgoudis for his early work on the massive
double box's UT basis. We acknowledge Dmitry Chicherin, Johannes Henn,
David Kosower, Pascal Wasser, and Li Lin Yang for very useful discussions.

The work of XL was supported from Qiu-Shi Funding and the National Natural Science Foundation
of China (NSFC) with Grant No.11935013, No.11575156. The work of YZ was supported from the NSF of China through Grant
No. 11947301, 12047502 and No. 12075234.

\appendix

\section{The perfect square and Gram-determinant relations}
\label{sec:Gramrels}
 To derive eq.\ \eqref{eq:gramder}, we use the following relation between Gram-determinants:
\begin{equation}
\label{eq:gram1}
 G(p_3,p_4,k_1)G(p_2,p_3,p_4)-G(p_2,p_3,p_4,k_1)G(p_3,p_4)=G(\{p_2,p_3,p_4\},\{p_3,p_4,k_1\})^2
\end{equation}
or
\begin{equation}
 u_3\Delta+u_2\lambda_4=u_4^2,
\end{equation}
with $u_4=G(\{p_2,p_3,p_4\},\{p_3,p_4,k_1\})$ and $\lambda_4=(m_3^2-s)^2$.
When substituting $x_3\rightarrow c_3^\pm$, the first term on the r.h.s.\ vanishes and therefore
\begin{equation}
\label{eq:quadr}
 u_{2,3^\pm}\lambda_4=u_{4,3^\pm}^2.
\end{equation}

We can simplify even more by using
\begin{equation}
 G(\{p_2,p_3,p_4\},\{p_3,p_4,k_1\})=-\frac{1}{2}\frac{\partial}{\partial (2k_1\cdot p_2)}G(p_2,p_3,p_4,k_1)
\end{equation}
or
\begin{equation}
\label{eq:g4der}
 u_4=\frac{1}{2}\frac{\partial}{\partial z_2}u_2.
\end{equation}

The relation between the Gram determinants in \eqref{eq:gram1} can be generalized to\footnote{We thank Li Lin Yang for sharing this identity with us.}
\begin{align}
\label{eq:gramGen}
 G(k,\vec{q})G(p,\vec{q})-G(k,p,\vec{q})G(\vec{q})=G(\{k,\vec{q}\},\{p,\vec{q}\})^2,
\end{align}
where $\vec{q}$ can be any sequence of vectors. Further, we have
\begin{equation}
\label{eq:Gramder}
 G(\{k,\vec{q}\},\{p,\vec{q}\})=-\frac{1}{2}\frac{\partial}{\partial(2k\cdot p)}G(k,p,\vec{q}).
\end{equation}
Relation \eqref{eq:gramGen} follows from
Sylvester's determinant identity\footnote{We acknowledge Dmitry Chicherin for making this observation.}
and relation \eqref{eq:Gramder} is just Jacobi's formula.

Back to eqs.\ \eqref{eq:quadr} and \eqref{eq:g4der}, we can use that $u_4$ is linear in the $x_i$ and therefore
\begin{equation}
 u_{4,3^\pm}=\frac{1}{2}\left[1-(x_3-c_3^\pm)\frac{\partial}{\partial x_3}\right]\frac{\partial}{\partial z_2}u_2.
\end{equation}
Putting everything together shows that
\begin{equation}
 \sqrt{u_{2,3^\pm}}=\frac{1}{\sqrt{\lambda_4}}\frac{1}{2}\left[1-(x_3-c_3^\pm)\frac{\partial}{\partial x_3}\right]\frac{\partial}{\partial z_2}u_2.
\end{equation}


\bibliographystyle{JHEP}
\bibliography{bibtex}

\end{document}